\documentclass[aps,prd,twocolumn,noshowpacs,superscriptaddress,groupedaddress,nofootinbib,preprintnumbers]{revtex4}

\usepackage{amsmath}
\usepackage{amsfonts}
\usepackage{amssymb}
\usepackage{textcomp}
\usepackage{graphicx, rotating}
\usepackage{epstopdf}
\usepackage{epsfig}
\usepackage{latexsym}
\usepackage{graphicx}
\usepackage{color}
\usepackage{amsmath,amssymb}
\usepackage{slashed}
\usepackage{hyperref}
\usepackage{multirow}

\hypersetup{colorlinks, citecolor=bluscuro, linkcolor=black, urlcolor=bluscuro}
\definecolor{rossos}{cmyk}{0,1,1,0.55}
\definecolor{bluscuro}{rgb}{0.15, 0.2, .85}
\definecolor{bluchiaro}{cmyk}{1,.3,0.,0.1}
\definecolor{green1}{rgb}{0.0, 0.5, .0}

\newcommand{\eq}[1]{Eq.~(\ref{#1})}

\newcommand{\pslash}{\!\not\! p}

\newcommand{\be}{\begin{equation}}
\newcommand{\ee}{\end{equation}}
\newcommand{\bea}{\begin{eqnarray}}
\newcommand{\eea}{\end{eqnarray}}
\newcommand{\bc}{\begin{center}}
\newcommand{\ec}{\end{center}}

\newcommand{\ar}[1]{| {#1} \rangle}
\newcommand{\al}[1]{\langle {#1} |}

\newcommand{\dfour}{$D\,$$=\,$4 }
\newcommand{\dsix}{$D\,$$=\,$6 }
\newcommand{\deight}{$D\,$$=\,$8 }

\def\tr{{\rm tr\,}}
 \def\ra {\rightarrow}

 \def\op{{\cal O}}
\def\lra#1{\overset{\text{\scriptsize$\leftrightarrow$}}{#1}}

\definecolor{red}{rgb}{1,0,0}
\definecolor{pink}{rgb}{1,0,1}

\begin{document}

\title{Helicity Selection Rules and Non-Interference for BSM Amplitudes}

\author{Aleksandr~Azatov}
\affiliation{Abdus Salam International Centre for Theoretical Physics,
Strada Costiera 11, 34151, Trieste, Italy
}

\author{Roberto~Contino}
\thanks{On leave from Dipartimento di Fisica, Universit\`a di Roma “La Sapienza” and INFN, Roma, Italy.}
\affiliation{Institut de Th\'eorie des Ph\'enomenes Physiques, EPFL, Lausanne, Switzerland}
\affiliation{Theoretical Physics Department, CERN, Geneva, Switzerland}

\author{Camila~S.~Machado}
\affiliation{Instituto de F\'isica Te\'{o}rica, Universidade Estadual Paulista, SP, Brazil}
\affiliation{Theoretical Physics Department, CERN, Geneva, Switzerland}

\author{Francesco~Riva}
\affiliation{Theoretical Physics Department, CERN, Geneva, Switzerland}

\date{\today}

\begin{abstract}
Precision studies of scattering processes at colliders provide powerful indirect constraints on new physics.
We study the helicity structure of scattering amplitudes in the SM and in the context of an effective Lagrangian description of BSM dynamics.
Our analysis reveals a novel set of helicity selection rules according to which, in the majority of $2\to2$ scattering processes at high energy, the SM and the leading 
BSM effects do not interfere. In such situations, the naive expectation that dimension-6 operators represent the leading BSM contribution is compromised,
as corrections from dimension-8 operators can become equally (if not more) important well within the validity of the effective field theory approach.
\end{abstract}

\preprint{CERN-TH-2016-165}

\maketitle

\section{Introduction}\label{sec:intro}

\enlargethispage{0.5cm}
Standard Model (SM) precision tests represent an important strategy in the search for new physics.
Effective field theories provide a suitable theoretical framework which allows these tests to be performed model-independently while maintaining
a simple connection to explicit UV theories.  The effective field theory (EFT) approach is 
especially convenient to organize hierarchically possible departures from the SM. 
Models  in which a large separation exists between the new physics scale $\Lambda$ and the electroweak (EW) scale 
can be expanded in powers of fields and derivatives~\footnote{In the following we will assume for simplicity that the UV dynamics conserves baryon and 
lepton numbers.}
\begin{equation}\label{lag}
{\cal L}={\cal L}^{\rm SM}+{\cal L}^{6}+{\cal L}^{8}+\cdots,\quad {\cal L}^{D}=\sum_i c^{(D)}_i\op^{(D)}_i,
\end{equation}
where $c^{(D)}_i\sim \Lambda^{4-D}$ and $D$ is the dimension of the operator $\op^{(D)}_i$.
In most theories,  \dsix terms are expected to capture the leading beyond-the-SM (BSM) effects. (In the presence of approximate symmetries or other selection rules, 
effects from \dsix operators can be suppressed compared to those from \deight or higher-dimensional operators, see ~\cite{rattetaltoappear,eftvalidity}.)
This motivates searches for generic new physics, as parametrized by ${\cal L}^{6}$ only~\cite{Aad:2015eha,Aad:2015tna,CMS:2015uda}. 
In particular, when departures from the SM are small, as typically occurs in weakly-coupled theories, the leading corrections to the cross section are expected to arise 
at order  $1/\Lambda^2$ from the interference between the SM and \dsix operators.
Aim of this note is to assess the validity of this naive expectation by analyzing the relative importance of the contributions to scattering amplitudes 
from the different terms in \eq{lag}.

Precision searches can be divided into two categories: \textit{i)} those exploiting the resonantly enhanced production of a SM state (such as measurements at the 
$Z$-pole or single-Higgs production); \textit{ii)} those exploiting the high-energy $E\gg m_W$ behavior of non-resonant processes (including  $e^+e^-\to f\bar f$ 
at LEP2 and $W^+W^-$  production).  This second mode of exploration is ubiquitous in the LHC experimental 
program~\cite{Aad:2015eha,CMS:2015uda,CMS:2015jaa,CMS:2015ibc,Aad:2014zda}, as an obvious consequence of its high-energy reach, and it will be the focus 
of this work.

We anticipate our main result in Table~\ref{tab:hel}: in the high-energy (massless) limit and working at tree level, 
SM and \dsix BSM contributions to $2\to 2$ scattering processes involving 
at least one transversely-polarized vector boson appear in different helicity amplitudes and thus do not interfere. 
This non-interference rule contradicts the naive expectation and implies that in these processes
\dsix and \deight operators contribute at the same order in the $1/\Lambda$ expansion if masses and loop corrections are neglected.
It follows that in many cases of interest analyses based on an EFT truncated at the $D=6$ level are incomplete in the high-energy region away from threshold.
\begin{table}[hbt]
\begin{center}
\begin{tabular}{|c|c|c|}
\hline
$A_4$ & $|h(A_4^{\rm SM})|$ &$|h(A_4^{\rm BSM})|$\\
\hline
$VVVV$ & 0 & 4,2 \\
\hline
$VV\phi\phi$ & 0 & 2 \\
\hline
$VV\psi\psi$ & 0 & 2 \\
\hline
$V\psi\psi\phi$ & 0 & 2 \\
\hline
\hline
$\psi\psi\psi\psi$ & 2,0 & 2,0 \\
\hline
$\psi\psi\phi\phi$ & 0& 0 \\
\hline
$\phi\phi\phi\phi$ & 0 & 0 \\
\hline
\end{tabular}
\end{center}
\caption{\emph{Four-point amplitudes $A_4$ {that do not vanish in the massless limit and the total helicity $h(A_4)$ of their SM and BSM contributions.} \mbox{$V=V^\pm$}, $\psi=\psi^\pm$ and $\phi$  denote, respectively, transversely-polarized vectors, fermions (or antifermions) and scalars in the SM. For processes with at least one transversely-polarized vector (listed above the double line in the table), SM and BSM contributions do not interfere in the massless limit because have different total helicity. }}\label{tab:hel}
\end{table}%

\section{Helicity Selection Rules and Non-Interference}
\label{sec:helicities}

When departures from the SM are small, the leading BSM contribution comes from the SM-BSM interference term in the amplitude squared. Obviously, interference is possible only if SM and BSM give non-vanishing contribution to the same helicity amplitude. In this section we study the helicity structure of scattering amplitudes at tree-level, in the SM and at  leading  order in the effective field theory expansion, i.e. at the level of \dsix operators.
We will denote the corresponding new-physics contribution as BSM$_6$ in the following.  We focus first on the phenomenologically relevant case of $2\to2$ scatterings 
and work in the massless limit; the massive case and higher-points amplitudes are discussed below. We use the spinor-helicity formalism (see
Refs.~\cite{Dixon:1996wi,Elvang:2013cua} for a review), where the fundamental 
objects which define the
scattering amplitudes are Weyl spinors $\psi^\alpha$ and $\bar\psi_{\dot \alpha}$, transforming as $(1/2,0)$ (undotted indices) and $(0,1/2)$ (dotted indices) representations of $SU(2)\times SU(2)\simeq SO(3,1)$, and Lorentz vectors $A_\mu\sigma^\mu_{\alpha\dot \alpha}$, transforming as $(1/2,1/2)$.~\footnote{We will not distinguish between fermions and anti-fermions except where explicitly 
mentioned,  as this distinction is not crucial to our analysis. We will denote a Weyl fermion or anti-fermion of helicity $+$ ($-$) with $\psi^+$ ($\psi^-$). When
indicating a scattering amplitude, the symbol $\psi$ will stand for either $\psi^+$ or $\psi^-$.} 
In this language, the field strength is written as
\begin{equation}
F_{\mu\nu}\sigma^\mu_{\alpha\dot \alpha}\sigma^\nu_{\beta\dot\beta}
\equiv F_{\alpha\beta}\bar \epsilon_{\dot{\alpha}\dot{\beta}}+\bar F_{\dot\alpha\dot\beta} \epsilon_{\alpha \beta}
\end{equation}
in terms of its self-dual and anti-self dual parts $F$ and $\bar F$ (transforming respectively as $(1,0)$ and $(0,1)$ representations).

\begin{figure}[tb]
\begin{center}
\begin{picture}(155,55)
\put(0,0){\includegraphics[width=0.3\textwidth]{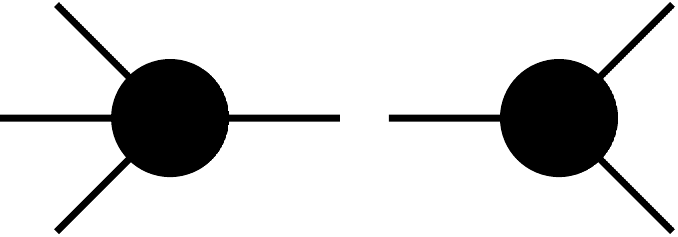} }
\put(28,41){$A_m$}
\put(110,41){$A_{m'}$}
\put(60,30){$\pm$}
\put(85,30){$\mp$}
\end{picture}
\end{center}
     \caption{\emph{When the factorization channel goes on-shell, it propagates a well-defined helicity eigenstate and \eq{sumH} holds.}}\label{fig:facto}
\end{figure}
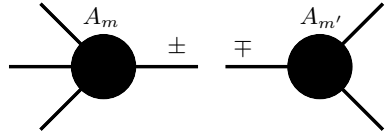

Our  analysis will be in terms of complex momenta $p\in\mathbb{C}$: 
this allows one to make sense of 3-point amplitudes on-shell, even though these vanish for massless states with real kinematics.
We will need three well-known results, that we summarize here and discuss in the Appendices, see Refs.~\cite{Dixon:1996wi,Elvang:2013cua,Cohen:2010mi}.  
These are:

\vspace{0.2cm} \noindent
{\bf 1.} \hspace{0.05cm}  Consider an amplitude $A_n$ with $n$ external legs ($n$-point amplitude), and let $A_m$ and $A_{m'}$ be any
two sub-amplitudes,  with $m+m'-2=n$, see Fig.~\ref{fig:facto}.
%
%
We define the net helicity of an on-shell amplitude, $h(A)$, as the sum of the helicities of all its external states, where all momenta are taken to be outgoing.
Then one has:
\begin{equation}\label{sumH}
h(A_n)=h(A_m)+h(A_{m'})
\end{equation}
for all possible sub-amplitudes $A_m$ and $A_{m'}$.
This relation is a consequence of the fact that the amplitude has a pole when the  intermediate line goes on-shell, and that in this limit it
factorizes into the product of the two sub-amplitudes. 
While in the SM there are no exceptions to Eq.~(\ref{sumH}), in the \dsix effective theory this relation fails if
an effective operator gives a contribution to the vertex attached to the intermediate line that vanishes on shell.
In this case the pole from the propagator is canceled by the vertex, and factorization does not hold.
When this occurs  the operator can be rewritten through the equations of motion in terms of other operators with more fields.
We will discuss below how this complication is avoided.

\vspace{0.2cm} \noindent
{\bf 2.} \hspace{0.05cm} Dimensional analysis, Little group scaling and the 3-particle special kinematics fix completely the form of the 3-point amplitudes,
and in particular relate their total helicity $h(A_3)$ to the dimensionality of the coupling~$g$ characterizing the 3-point vertex:
\begin{equation}\label{sumh}
|h(A_3)|=1-\lbrack g \rbrack\,.
\end{equation}
For instance, the triple gauge interaction of the SM is characterized by a dimensionless coupling, and the corresponding
3-point on-shell amplitude has $|h|=1$.
The \dsix operator ${\cal O}_{3W}=\tr\!(W_{\mu\nu}W^{\nu}_\rho W^{\rho\mu})$ instead appears in \eq{lag} with a coefficient $c_{3W}$ with dimension $[c_{3W}]=-2$,
and thus generates a 3-point amplitude with $|h|=3$.

\vspace{0.2cm} \noindent
{\bf 3.} \hspace{0.05cm}
Helicity selection rules in the SM force the following 4-point amplitudes with $|h|=2$ to vanish:
\begin{multline} \label{MHV}
A(V^+V^+V^+V^-)=\,A(V^+V^+\psi^+\psi^-)\\[0.1cm]
=A(V^+ V^+\phi\phi)=A(V^+\psi^+\psi^+ \phi)=0\, .
\end{multline}
These relations can  be proved by means of the Supersymmetric Ward Identities (SWI) \cite{Grisaru:1976vm,Parke:1985pn}, as sketched in Appendix~\ref{a:susy}.
In the limit in which all up-type or all down-type Yukawa couplings vanish,
the SM Lagrangian can be uplifted to a supersymmetric one,
for which SWI hold. Such theory has in addition an $R$-parity implying that the supersymmetric partners do not contribute at tree level to scattering amplitudes 
with external SM legs only. As a consequence, \eq{MHV} holds  for the SM.

\subsection{The Standard Model}
\label{sec:SM}

Within the SM, it follows from property {\bf 2} that 3-point vertices associated with marginal couplings have
\begin{equation}\label{H3SM}
 h(A_3^{\rm SM})=\pm 1 \, .
\end{equation}
The three-scalar vertex (which would have vanishing total helicity) is absent in the SM in the massless limit (i.e. in the limit of unbroken EW symmetry).
With the exception of the quartic scalar vertex, which has trivially total helicity $h(A_4)=0$,  all 4-point on-shell vertices in the SM can be made vanish with 
a suitable definition of polarization vectors (this is a direct consequence of gauge invariance, see for example~\cite{Elvang:2013cua}).
Properties~{\bf 1} and~{\bf 3} then imply that all 4-point amplitudes with fermions or gauge fields have vanishing total helicity, unless they involve both up- and down-type 
Yukawa interactions~\cite{Parke:1986gb,Berends:1987me,Dixon:1996wi}.
The only exceptions are in fact the amplitudes $\psi^+\psi^+\psi^+\psi^+$ and $\psi^-\psi^-\psi^-\psi^-$, which receive a contribution 
(proportional to the product of up- and down-type Yukawas)
from the Higgs exchange and have $|h(A_4)|=2$.
These known results are summarized in Table~\ref{tab:hel}.

\subsection{Beyond the Standard Model}
\label{sec:BSM}
{Local operators entering at a given order  in the $1/\Lambda$ expansion of an EFT  
can be redefined by making use of the equations
of motion (EoM) derived at lower order.
For example, it is always possible to rewrite \dsix operators by using the EoM of the renormalizable $D$=4 Lagrangian; the new effective
Lagrangian will differ from the original one by \deight terms.
This freedom allows one to systematically replace operators with more derivatives in terms of operators  involving more fields.
At the \dsix level, this procedure leads to the so called Warsaw basis of operators introduced in Ref.~\cite{Grzadkowski:2010es}.}
This basis is particularly convenient to study  $2\ra 2$ scattering processes for two reasons.
First, the number of operators with two and three fields (bivalent and trivalent operators) is reduced to a minimum. In particular,
there are no bivalent operator and only two trivalent ones, i.e.  $\op_{3W}$ and $\op_{3\widetilde W}=\tr\!(W_{\mu\nu}W^{\nu}_\rho \widetilde W^{\rho\mu})$; 
all the other \dsix operators have at least four fields, see Table~\ref{t: helicityop}. 
A second, more important, reason why the Warsaw basis is convenient is because condition {\bf 2} requires that sub-amplitudes do not 
receive contributions that are vanishing on shell (but different from zero off shell). 
Such contributions would be proportional to inverse SM propagators and thus
arise from local operators that vanish on the \dfour EoM.
Eliminating higher-derivative redundant operators proportional to the EoM automatically guarantees that the amplitudes factorize and 
\eq{sumH} is  fulfilled.
As an example, consider the operator ${\cal O}_B=(i/2)H^\dagger  \lra {D^\mu} H \,\,\partial^\nu  B_{\mu \nu}$, which appears in the SILH basis of 
Ref.~\cite{Giudice:2007fh}. It gives a vanishing contribution to the on-shell $HHB$ vertex (even for complex momenta), but contributes off shell to processes 
like $HH\to HH$ or $HH\to \psi\psi$. Indeed, by using the EoM it can be eliminated in favor of operators of the form $D^2H^4$ or $H^2D\psi^2$, 
that contribute to the previous processes via contact interactions.

In order to determine the helicity of an amplitude generated through the insertion of some operator ${\cal O}$, it is useful to introduce 
the holomorphic and anti-holomorphic weights of ${\cal O}$, as defined by Ref.~\cite{Cheung:2015aba}. 
For an arbitrary on-shell amplitude $A$ with $n(A)$ legs and helicity $h(A)$,
\bea
\label{eq:wdef}
w(A)= n(A)-h(A), \quad \bar {w}(A)= n(A)+h(A).
\eea
The weights of the operator ${\cal O}$ are then obtained by minimizing over all the amplitudes involving ${\cal O}$:
\begin{equation} \label{eq:wOdef}
w({\cal O})=\min_{A} \{w(A)\}\, , \quad
\bar w({\cal O})=\min_{A} \{\bar w(A)\} \, .
\end{equation}
The point is that, as a consequence of \eq{sumH} and the fact that $h(A^{\rm SM}_3)= \pm 1$, 
building amplitudes
with more SM interactions cannot 
decrease $w(A)$ and $\bar w(A)$, so that the weight is always determined 
by the amplitude with the smallest number of SM vertices.
Since they are defined in terms of
on-shell amplitudes, weights offer various advantages. First of all, they are gauge-invariant quantities  
characterizing also operators, involving covariant derivatives, that contribute to different processes with different helicities. 
Moreover, they are well defined even for operators whose contribution to a given amplitude vanishes 
on shell (as for~${\cal O}_B$ discussed above). Finally, one can easily deduce from Eqs.~(\ref{eq:wdef}) and~(\ref{eq:wOdef}) 
that the helicity of $n$-point amplitudes with one~${\cal O}$ insertion is constrained to be in the range
\bea
\label{eq:w}
\bar w ({\cal O})-n\leq h(A_n^{\cal O})\leq n-w ({\cal O})\, .
\eea

Using these ingredients we can  constrain the total helicity of BSM$_6$ contributions to $2\to 2$ scattering amplitudes. Let us start with the unique trivalent 
$F^3$ and $\bar F^3$ structures of ${\cal O}_{3W}$ and ${\cal O}_{3\widetilde W}$. 
Given the dimensionality of their coefficients, $[c_i]=-2$, Eq.~(\ref{sumh}) fixes the helicity of their contribution to the 3-point amplitude up to a sign: 
$|h(A_3)|=3$. It is in fact not difficult to show that $F_{\alpha\beta}$ and $\bar F_{\dot\alpha\dot\beta}$ generate states with helicity $+1$ and $-1$ respectively 
(see~\cite{Witten:2003nn,Dixon:2004za}), which 
implies that $h(A_3)= +3$ for $F^3$ and $h(A_3)= -3$ for $\bar F^3$. From Eq.~(\ref{eq:wOdef}) it follows that the weights of $F^3$ and $\bar F^3$
are respectively $(w,\bar w)=(0,6)$ and $(w,\bar w)=(6,0)$.
Equation~(\ref{eq:w}) thus constrains the helicity of a 4-point function with one insertion of either of these operators to be in the range $2\leq |h(A_4)|\leq 4$
(more precisely, $2\leq h(A_4)\leq 4$ for $F^3$ and $-4\leq h(A_4)\leq -2$ for $\bar F^3$).
Considering that $h(A_4)=0$ in the SM (for the amplitudes under consideration), this shows that no SM-BSM interference is possible in this case.
It is useful to directly verify the constraint of \eq{eq:w}  for some specific amplitudes.
Starting from a 3-point amplitude with one $F^3$ insertion, for example, a 4-point one is obtained by adding a SM cubic vertex (which has $h(A_3^{\rm SM})=\pm1$ 
as shown previously). Then \eq{sumH} implies that the 4-point amplitude with only vectors, $VVVV$, has $|h|= 4, 2$ \cite{Dixon:1993xd,Dixon:2004za} 
(notice that  $F^3$ is not supersymmetrizable and condition {\bf 3} does not  hold \cite{Cohen:2010mi,Broedel:2012rc}). Similarly, the helicity of an amplitude 
$VV\psi\psi$,  is  $|h|= 2$. Both results agree with the bound of \eq{eq:w}.

\begin{table}[]
\centering
\begin{tabular}{l@{\hskip 0.1in}c@{\hskip 0.1in}c@{\hskip 0.1in}c@{\hskip 0.1in}c}
\hline
$\mathcal{O}_i$&$n_{min}$ & $h_{min}$ &$(w,\bar w)$ &  $c_i$   \\[0.1cm]
 \hline
$F^3$    &3& 3 & (0,6)         &    $g_*/\Lambda^2$ \\[0.1cm]
$F^2\phi^2$, $F\psi^2\phi$,  $\psi^4$    &4& 2 & (2,6)          &   $g_*^2/\Lambda^2$    \\[0.1cm]
$\psi^2\bar\psi^2$, $\psi\bar\psi\phi^2 D$, $\phi^4 D^2$ &4&  0 & (4,4)                &  $g_*^2/\Lambda^2$ \\[0.1cm]
$\psi^2\phi^3$ &5& 1& (4,6)             &  $g_*^3/\Lambda^2$  \\[0.1cm]
$\phi^6$ &6 & 0& (6,6)	&$g_*^4/\Lambda^2$ \\[0.1cm]
\hline
\end{tabular}
\caption{\emph{Weights  $(w,\bar w)$ of the dimension-6 operators ${\cal O}_i$ in the Warsaw basis.
Also shown are the number of legs $n_{min}$ and corresponding helicity $h_{min}$ of the smallest amplitude to which the operator
contributes, and the naive estimate of its coefficient $c_i$.
Operators with $\psi_\alpha\leftrightarrow{\bar\psi_{\dot\alpha}}$ and $F_{\alpha\beta}\leftrightarrow\bar{F}_{\dot\alpha\dot\beta}$  have $h_i\to-h_i$  hence $(w,\bar w) \leftrightarrow ( \bar w,w)$.}}
\label{t: helicityop}
\end{table}

Apart from $F^3$ and $\bar F^3$, the remaining \dsix operators of the Warsaw basis do not contribute to 3-point amplitudes in the massless (high-energy) limit. 
Those contributing to 4-point amplitudes are listed in the second and third row of Table~\ref{t: helicityop}. The helicity of the 4-point amplitudes in this case 
can be directly determined from the corresponding operators by noticing that $F_{\alpha\beta}$ ($\bar F_{\dot\alpha\dot\beta}$) creates states with helicity $+1$ $(-1)$,  
$\psi_{\alpha}$ ($\psi_{\dot \alpha}$) creates fermions or antifermions with helicity $+1/2$ $(-1/2)$, and the helicity of scalars trivially vanishes.
For example, an operator $F\psi^2\phi$ can excite states with net helicity $h_{min} =+2$, 
which equals the helicity of the corresponding 4-point amplitude.
The results are reported in Table~\ref{t: helicityop}: the operators in the second row lead to 4-point amplitudes with helicity 
$|h(A_4)|=2$, and thus do not interfere with the SM. The operators in the third row give $|h(A_4)|=0$ and can thus interfere with the SM, but the corresponding
amplitudes do not involve transversely-polarized vector bosons. These results directly imply those of Table~\ref{tab:hel}.

In addition to the helicity selection rules derived above, $2\to 2$ tree-level scattering amplitudes are constrained by additional selection rules in the massless limit.
In particular, a simple one follows from weak isospin conservation. 
In the Warsaw basis, the only BSM contribution to the amplitudes $VV\psi\psi$, $VVVV$ comes from $F^3$, $\bar F^3$, and it can always be written as the product of
two 3-point amplitudes with a vector propagator ($VV\psi\psi$ receives no quartic contribution from \dsix operators, while the quartic $VVVV$ vertex 
can always be made vanish through a suitable choice of polarization vectors). 
The propagation of a vector boson  implies a well defined $SU(2)_L$ isospin structure of the external states produced at each vertex: they 
transform in the  ${\bf 3}\in {\bf 3}\otimes {\bf 3}$, which is totally antisymmetric and thus does not include pairs of identical bosons.
For this reason amplitudes like $ZZZZ$, $\gamma\gamma\gamma\gamma$, $\psi\psi ZZ$ and $\psi\psi\gamma\gamma$ can only be generated by
\deight operators.
It is worth mentioning another selection rule which characterizes the \dsix effective theory in the massless limit. 
Its Lagrangian is invariant under the $Z_2$ chiral symmetry 
\begin{equation}
\label{eq:chiralinv}
\phi\to-\phi, \quad \psi_L\to-\psi_L, \quad \psi_R\to +\psi_R\, ,
\end{equation}
as a direct consequence of $SU(2)_L$ invariance and of the SM quantum numbers (it is not possible to form operators which are singlets of $SU(2)_L$ with an 
odd number of $\psi_L$ and $H$ fields). It follows that the amplitudes $VVV\phi$ and $V\phi\phi\phi$ identically vanish (in the massless limit), while 
the helicities of the fermion and anti-fermion in  $V\psi\psi\phi$  are forced to be the same.
Notice that these same conclusions are also a consequence of
the helicity selection rules, since by the arguments presented above 
no 4-point amplitude has total helicity $|h|=1,3$ in the SM or at the \dsix level.
Using Eq.~(\ref{eq:chiralinv}) might still be useful, however, as a quicker way to determine if a given amplitude vanishes, independently of helicity arguments.

To summarize, we have shown that working at tree level and in the massless (i.e. high-energy) limit, the BSM contribution never interferes with the SM one in 
$2\to 2$ scattering amplitudes involving at least one transversely-polarized vector boson. Interference is possible, instead, for amplitudes involving only scalars 
(including longitudinally-polarized
vector bosons) and fermions, such as $\psi\psi\to\psi\psi$, $\psi\psi\to\phi\phi$ and $\phi\phi\to\phi\phi$.
We will comment on the practical implications of these results in Section \ref{sec:implications}, but first we discuss how our analysis generalizes to the
massive case and to higher-point scattering amplitudes.

\subsection{Higher-point Amplitudes}
\label{sec:higherpoint}

The helicity of amplitudes with 5 or more external legs can be easily determined in the SM by starting from that of 4-point amplitudes, given
in Table~\ref{tab:hel}, using the addition rule (\ref{sumH}) and knowing that 3-point vertices change helicity by $\pm 1$ unit (Eq.~(\ref{H3SM})).
We find
\begin{equation}\label{ASM}
\begin{gathered}
|h(A_{n\geq5}^{\rm SM})|\leq n-4\phantom{.} \\[0.1cm]
\textrm{with $h$ even (odd) for $n$ even (odd)}.
\end{gathered}
\end{equation}
For example, one has $h(A^{SM}_5) = \pm 1$, $h(A^{SM}_6) = 0, \pm 2$ respectively for 5-point and 6-point SM amplitudes.
By making use of  the helicity selection rules for 4-point amplitudes \eq{MHV}, combined with \eq{sumH}, one reproduces the well known result that
the first non-vanishing amplitudes with largest total helicity are the Maximal Helicity Violating ones~\cite{Dixon:1996wi,Elvang:2013cua}.

The helicity of BSM$_6$ amplitudes, including those with 5 or more external legs, can be constrained by using \eq{eq:w} and the weights reported 
in Table~\ref{t: helicityop}. We find:
\begin{equation}\label{ABSM}
\begin{gathered}
h_{min}^{\cal O} \leq |h(A_{n\geq n_{min}}^{\cal O})|\leq h_{max}^{\cal O} \\[0.1cm]
\textrm{with $h$ even (odd) for $n$ even (odd)}
\end{gathered}
\end{equation}
\begin{center}
\begin{tabular}{l@{\hskip 0.2in}c@{\hskip 0.2in}c}
${\cal O}_i$ & $h_{min}^{\cal O}$ & $h_{max}^{\cal O}$ \\[0.1cm]
\hline
$F^3$    & $6-n$ & $n$ \\[0.1cm]
$F^2\phi^2$, $F\psi^2\phi$,  $\psi^4$    & $6-n$ & $n-2$     \\[0.1cm]
$\psi^2\bar\psi^2$, $\psi\bar\psi\phi^2 D$, $\phi^4 D^2$ & 0 & $n-4$   \\[0.1cm]
$\psi^2\phi^3$ & $6-n$ & $n-4$  \\[0.1cm]
$\phi^6$ & $0$ & $n-6$  
\end{tabular}
\end{center}
where $n_{min}$ is given in Table~\ref{t: helicityop} for the various operators.

Notice that, both for the SM and BSM$_6$, the total helicity is even (odd) if $n$ is even (odd). This follows from Little group scaling and the
even dimensionality of the coupling constants ($[g]=0$ in the SM and $[c_i]=-2$ in BSM$_6$), see Appendix~\ref{a:def}.
In this respect (as well as to derive Eq.~(\ref{ASM})), it is crucial that no scalar cubic vertex is present in the SM in the limit of unbroken EW symmetry.
This selection rule automatically implies that amplitudes such as $VVV\phi$ or $VV\phi\phi\phi$ must vanish since they would have necessarily 
a total helicity with the wrong parity (a similar conclusion follows from chiral invariance, as seen above).

From Eq.~(\ref{ABSM}) it follows that all \dsix operators contribute to amplitudes with $h(A_5)=\pm 1$ or $h(A_6)=0,\pm 2$ (the operator $\phi^6$ 
contributes only to 6-point amplitudes), and can thus potentially interfere with the SM.
Having the same total helicity is in fact a necessary but not sufficient condition for the SM and BSM amplitudes to have interference. 
The same net helicity can indeed be distributed differently on the external legs, in which case no interference occurs. 
As an example consider the 5-point amplitudes $\psi^-\psi^- \phi V^+ g^+$ and $\psi^+\psi^+ \phi V^+ g^-$, both with $h=+1$,
where $\psi$ is a quark,  $g$ a gluon and $V=W,Z,\gamma$. The SM contributes only to the first as a consequence of the addition rule (\ref{sumH})
and the fact that 4-point sub-amplitudes have necessarily $h=0$. The second amplitude instead can be generated by the insertion of an 
EW dipole operator $F\psi^2\phi$, which forces the helicity of the vector boson to have the same sign  as that of the quarks.
Moreover, in the case in which both the SM and BSM$_6$ amplitudes have the same external helicities, additional selection rules can forbid the interference.
As an example, consider the amplitude $g^+g^+\psi^-\psi^-\phi$ (where $\psi$ could be a top quark, $\phi$ a Higgs boson
and the corresponding physical scattering $gg\to t\bar t H$), which is non-vanishing in the SM in the massless limit.
The operator ${\cal O}_{3\widetilde G}$ contributes to the same helicity amplitude, but no interference occurs because ${\cal O}_{3\widetilde G}$ is CP odd.
Except for these particular cases, however, \dsix operators will in general interfere with the SM at the level of $n\geq 5$ amplitudes.
Non-interference seems therefore a peculiarity of 4-point amplitudes.

\subsection{Finite-Mass Effects and Radiative Corrections}
\label{sec:massive}

The non-interference between SM and BSM$_6$ amplitudes holds for $2\to 2$ scatterings in the massless limit and at tree level. 
There are two main subleading effects which correct this result in real scattering processes: finite-mass corrections and radiative effects (1-loop corrections
and real emissions).

Finite-mass effects can be easily included in our analysis. They 
can be parametrized in terms of
\begin{equation}
\varepsilon_V \equiv \frac{m_V}{E}\,,\qquad \varepsilon_\psi \equiv \frac{m_\psi}{E}\,,
\end{equation}
where $m_{V(\psi)}$ is the vector (fermion) mass and $E$ the energy. 
Finite-mass effects have been extensively studied in the literature, see \cite{Boels:2011zz,Coradeschi:2012iu,Badger:2005jv,Badger:2005zh,Ozeren:2006ft}.
In this note we are interested, in particular, to determine at which order in $\varepsilon_{V,\psi}$ the leading correction to a given amplitude appears.
To this aim, the most effective procedure is to consider higher-point amplitudes with additional Higgs bosons in the external legs.
Restricting to the phase space region where no momentum is transferred through a Higgs line corresponds in fact to setting the Higgs field to its vev.
One can thus identify the contribution to a given amplitude at order $k$ in the $\varepsilon$ expansion by considering a higher-point amplitude with
$k$ insertions of the Higgs. Since the  gauge and Yukawa interactions of the Higgs violate helicity by $\pm 1$ unit, this procedure allows one to easily
determine the leading contribution to the $n=4$ amplitudes that are vanishing in the massless limit.
For example, a transversely polarized vector can be turned into a longitudinal one (or vice versa) at order $\varepsilon_V$ 
through the insertion of a vertex $\phi^*\partial_{\mu}\phi A^{\mu}$, by setting $\phi^*$ to its vev. 
This follows from the Equivalence Theorem~\cite{Chanowitz:1985hj,Wulzer:2013mza}, which states that, at leading order in the $\varepsilon_V$ expansion, 
a longitudinal polarization can be replaced by the corresponding would-be Nambu-Goldstone boson.
Notice that it is not possible to \textit{flip} the helicity of a given vector line by inserting two Higgs vevs (e.g. 
making use of  two consecutive $\phi^*\partial_{\mu}\phi A^{\mu}$ interactions or inserting one $\phi^*\phi A^\mu A_\mu$ vertex), since the $VV\phi\phi$ 
sub-amplitude has vanishing total helicity in the SM.
For fermions, the Yukawa interaction $\psi^\alpha\psi_\alpha\phi$ ($\psi_{\dot\alpha}\psi^{\dot\alpha}\phi$)  has total helicity $h=+1$ $(-1)$ and its insertion 
leads to a flip of the fermion chirality at order $\varepsilon_{\psi}$, according to \eq{sumH}.

In general, the SM-BSM$_6$ interference in an $n=4$ amplitude will arise at order $k_{SM}+k_{BSM}$ in $\varepsilon$
if non-vanishing SM and BSM$_6$ amplitudes exist with respectively $k_{SM}$ and $k_{BSM}$ additional external Higgs fields.
The power $k_{SM}+k_{BSM}$ is always even, as a consequence of the fact that the total helicity of an $n$-point amplitude is even (odd) if $n$ is even (odd).
Hence, the interference in $n=4$ scattering amplitudes with one or more transversely polarized vector bosons is suppressed at least by two powers 
of $\varepsilon$.
As an example of the case $k_{SM} = 0$, $k_{BSM} =2$ consider the amplitude $V^+V^+V^-V^-$:  it is non-zero in the SM but its
BSM$_6$ contribution vanishes in the massless limit. The 6-point amplitude $V^+V^+V^-V^-\phi\phi$, on the other hand, is generated by  $F^3$ and $F^2\phi^2$.
At order $\varepsilon_V^2$ this leads to a contribution to the $n=4$ amplitude and thus to an interference with the SM.
As a second example, the amplitude $V^+V^+V^+V^+$ has interference at order $\varepsilon_V^4$ for $k_{SM} =4$, $k_{BSM} =0$.
The SM contribution arises from $V^+V^+V^+V^+\phi\phi\phi\phi$ after taking four Higgs vevs, while the BSM one is generated already in the massless limit
from insertions of~$F^3$.
Finally, an example with $k_{SM} =1$, $k_{BSM} =1$ is given by the amplitude $V^+\psi^+\psi^-\phi$, which vanishes in the massless limit both in
the SM and beyond. The 5-point amplitude $V^+\psi^+\psi^-\phi\phi$, on the other hand, is generated through one insertion of any of the operators
$F^3$, $F^2 \phi^2$, $F \psi^2 \phi$ and $\psi\bar\psi\phi^2 D$, and is non-vanishing also in the SM. The interference arises at order $\varepsilon_V^2$
(in the case of $F^3$, $F^2 \phi^2$ and $F \psi^2 \phi$) and $\varepsilon_V\varepsilon_\psi$ (for $\psi\bar\psi\phi^2 D$).

Contributions from fermion mass insertions are  in general subdominant compared to those from vector mass insertions, with the exception of  processes 
involving the top-quark, for which $\varepsilon_\psi \approx \varepsilon_V$, like $bW\to th$~\cite{Farina:2012xp} and $tW\to tW$~\cite{Dror:2015nkp}.
In processes involving gluons instead of EW  vector bosons, top quark mass insertions are in fact the only way to get interference between SM and BSM$_6$.
An example is given by the scattering $gg\to t\bar t$~\cite{Cho:1994yu,Degrande:2010kt,Bramante:2014gda}, where the operators $F^3$ and $F\psi^2\phi$ 
(where $F$ is a gluonic field strength) interfere at order $\varepsilon_\psi^2$ with the SM.

We summarize our results for processes involving EW vector bosons in Table~\ref{t:WWhh}, where we report the order in $\varepsilon_V$ at which a given helicity 
amplitude appears, in the SM and BSM$_6$. For simplicity we work in the limit of vanishing Yukawa couplings 
and do not specify which \dsix operators give rise to an interference.
In all cases the interference term in the amplitude squared goes like a constant in the high-energy limit, $E\gg m_W$, except for the processes in the first line
of the two panels, where it scales as~$1/E^2$.

\begin{table}[tb]
\centering
{\raisebox{-4.mm}{\begin{tabular}{l@{\hskip 0.1in}c@{\hskip 0.1in}c}
\hline
Channel & SM & BSM$_6$  \\
 \hline && \\[-0.4cm]
$++++$ & $\varepsilon_V^4$ & $\varepsilon_V^0$ \\
$+++-$ & $\varepsilon_V^2$  & $\varepsilon_V^0$ \\
$++--$  & $\varepsilon_V^0$  & $\varepsilon_V^2$ \\[0.1cm]
\hline && \\[-0.4cm]
${\scriptsize +\frac{1}{2}}\, {\scriptsize -\frac{1}{2}} ++$ & $\varepsilon_V^2$ & $\varepsilon_V^0$ \\[0.05cm]
${\scriptsize +\frac{1}{2}}\, {\scriptsize -\frac{1}{2}}+-$  & $\varepsilon_V^0$ & $\varepsilon_V^2$ \\[0.05cm]
${\scriptsize +\frac{1}{2}}\, {\scriptsize -\frac{1}{2}} ~ 0 ~ +$ & $\varepsilon_V^1$ & $\varepsilon_V^1$ \\[0.05cm]
${\scriptsize +\frac{1}{2}}\, {\scriptsize -\frac{1}{2}} ~ 0 ~ 0$ & $\varepsilon_V^0$ & $\varepsilon_V^0$ \\[0.12cm]
\hline
\end{tabular}}}
\hspace{0.2cm}
\begin{tabular}{l@{\hskip 0.1in}c@{\hskip 0.1in}c}
\hline
Channel &SM &BSM$_6$      \\
 \hline && \\[-0.4cm]
$0+++$    &  $\varepsilon_V^3$    & $\varepsilon_V^1$ \\
$0++-$   &  $\varepsilon_V^1$    & $\varepsilon_V^1$\\[0.05cm]
$0~0++$  & $\varepsilon_V^2$    & $\varepsilon_V^0$ \\
$0~0+-$  & $\varepsilon_V^0$    & $\varepsilon_V^2$\\
$0~0~0~+$  & $\varepsilon_V^1$    & $\varepsilon_V^1$ \\
$0~0~0~0$ & $\varepsilon_V^0$    & $\varepsilon_V^{0}$ \\[0.08cm]
\hline
\end{tabular}
 \caption{\emph{Leading power of $\varepsilon_V$ at which a given helicity amplitude is generated in the SM and BSM$_6$. 
The first column indicates the process and the polarizations of the external states: 0 corresponds to a longitudinally-polarized vector boson or to a Higgs boson, 
$\pm$ to a transversely-polarized vector boson $V=W,Z,\gamma$, and ${\pm \frac{1}{2}}$ to a fermion.  
Yukawa couplings have been neglected for simplicity, and only non-vanishing amplitudes are shown.
Conjugate  amplitudes with $+\leftrightarrow -$ follow the same pattern. The $\varepsilon^0_V$ entries match those of Table~\ref{tab:hel}.
}}        \label{t:WWhh}
  \end{table}

Let us now consider radiative corrections. In general, 1-loop corrections to 4-point amplitudes violate the helicity selection rules discussed in Sections~\ref{sec:SM} 
and ~\ref{sec:BSM}, and thus generate a non-vanishing interference. The most relevant contribution arises from QCD corrections to amplitudes with external quarks 
or gluons. Pure EW loop corrections have a similar effect but are numerically smaller.
The emission of an extra gluon transforms a 4-point to a 5-point amplitude and can also lead to interference.
It is well known (see~\cite{Dixon:1996wi}) that in the limit in which one parton becomes soft, an $n$-point color-ordered amplitude
factorizes into the product of  the $(n-1)$-point amplitude made of the remaining hard partons times a singular eikonal factor.
Because of the helicity selection rules controlling 4-point amplitudes, it is thus clear that soft emissions of extra gluons or quarks cannot lead to interference; 
that is: the interference vanishes in the soft limit.
Similarly, when two  partons $i$ and $j$ become collinear, an $n$-point color-ordered amplitude factorizes into the $(n-1)$-point amplitude obtained by replacing
the $ij$ pair with an ``effective'' parton carrying its momentum, times a singular splitting function (see~\cite{Dixon:1996wi}).
Starting with a 5-point amplitude and taking the collinear limit, the selection rules acting on 4-point amplitudes force the helicity of the effective parton to be
opposite in the SM and BSM$_6$ cases. This implies that the SM-BSM$_6$ interference term, once integrated over the full phase space, is non-singular, that is:
the collinear singularity of the amplitude squared vanishes in the total cross section at order $1/\Lambda^2$~\cite{Dixon:1993xd}.
The absence of soft and collinear singularities in real emission processes  in turn implies that 
SM and BSM$_6$ amplitudes which vanish at tree level are IR-finite at 1-loop~\cite{Dixon:1993xd}.
Hence, although 1-loop QCD corrections and real emissions of extra gluons do lead to interference
between SM and BSM$_6$ amplitudes, no logarithmic enhancement is present in the collinear and soft limits. This means that the interference is suppressed
by a factor $\alpha_s/4\pi$, where $\alpha_s$ is evaluated at the high-energy scale characterizing the scattering process.

Summarizing, interference between SM and BSM$_6$  can arise in $2\to 2$ exclusive processes as a result of 1-loop corrections
and finite-mass effects, with a relative suppression of order, respectively, $\alpha_{s}/4\pi$ (or $\alpha_{em}/4\pi$ for processes without colored particles) and $m_W^2/E^2$.
Mass effects dominate at lower energies, while radiative corrections take over at energies $E \gtrsim m_W \sqrt{4\pi/\alpha_s}$. 
Similar conclusions hold in the case of processes where the final state is defined inclusively with respect to the emission of additional  QCD radiation.
In this case the leading SM-BSM$_6$ interference arises also from amplitudes with an additional gluon, while the pure SM contribution stems at lowest perturbative order.

A way to access the $1/\Lambda^2$  corrections from \dsix operators without any relative suppression of the signal compared to the SM irreducible 
background,  is instead considering exclusive $2\to 3$ scattering processes, where the additional particle could be a hard gluon. 
{In this case, as discussed in Section~\ref{sec:higherpoint}, the interference term arises at tree-level also in the massless limit, so that \emph{both} the SM and SM-BSM$_6$ interference are equally suppressed.
This strategy was for example proposed by the authors of
Ref.~\cite{Dixon:1993xd}, who suggested  to constraint the operator 
${\cal O}_{3G}$ using three-jet events to avoid the non-interference of \mbox{4-point} amplitudes already noticed  in~\cite{Simmons:1989zs}. 
For final states with one extra hard gluon, the gain in  signal/$\sqrt{\textrm{background}}$, compared to $2\to 2$ loop processes,
is only of order $\sqrt{4\pi/\alpha_s}$. Moreover, additional radiation does not necessarily  guarantee interference in the $2\to 3$ amplitude. For instance, in the specific case of the EW process $q\bar q\to VV, V\phi$  (with $V = W,Z,\gamma$),} simple inspection of the
tree-level Feynman diagrams further reveals that no interference arises by adding one extra gluon.~\footnote{In the massless limit, the amplitude $\psi\psi VVg$ receives an $|h|=3$ BSM$_6$ contribution from $F^3$ and an $|h|=1$ from the SM, so no interference is possible. For $\psi\psi V\phi g$, only $F\psi^2\phi$ contributes in BSM$_6$ but,  as noticed in the last paragraph of Section~\ref{sec:higherpoint}, $V$ and $g$ have always equal helicity in the SM and opposite  in BSM$_6$ amplitudes, so no interference.
} 
The emission of two gluons does seem to give interference, but in
that case the significance is further suppressed and there is no parametric gain compared to $2\to 2$ processes.
Despite the above considerations, studying $2\to 3$ processes seems in general a promising strategy  to constrain
\dsix operators, and deserves further investigation.

\section{Phenomenological Implications}\label{sec:implications}

The helicity selection rules of Section~\ref{sec:helicities} imply that the contributions from operators $F^3$, $F^2\phi^2$, $F\psi^2\phi$ and 
their conjugates never interfere with the SM ones in $2\to2$ scattering process at high energy and tree level, 
and that the interference is suppressed by powers 
of $(m_W/E)^2$ or $\alpha_s/4\pi$. What does this imply on the phenomenology at the colliders  ?

The impact of BSM precision searches performed using the EFT approach can be easily quantified in the context of theories  characterized by a single microscopic  
scale $\Lambda$ and a single (new) coupling $g_*$~\cite{Giudice:2007fh}. This  provides a power counting prescription to estimate the size of  the effective coefficients 
$c_i$ in \eq{lag} in terms of the  parameters of the UV theory:
\begin{equation}
\label{eq:counting}
c^{(D)}_i \sim \frac{g_*^{n_i-2}}{\Lambda^{D-4}}\, ,
\end{equation}
where $n_i$ counts the number of fields in $\op^{(D)}_i$. 
The corresponding estimate of the coefficient of each operator is reported in the last column of Table~\ref{t: helicityop}.
Such power counting smoothly interpolates between the strong coupling limit $g_*\sim 4\pi$, where it is equivalent to Naive Dimensional 
Analysis~\cite{Manohar:1983md}, and the weak coupling limit $g_*\lesssim g_{\rm SM}$. 
Additional symmetries and selection rules can lower the estimates of Eq.~(\ref{eq:counting})~\cite{Giudice:2007fh,rattetaltoappear,eftvalidity}.

\begin{figure*}[ht]
\begin{center}
\includegraphics[width=0.323\textwidth]{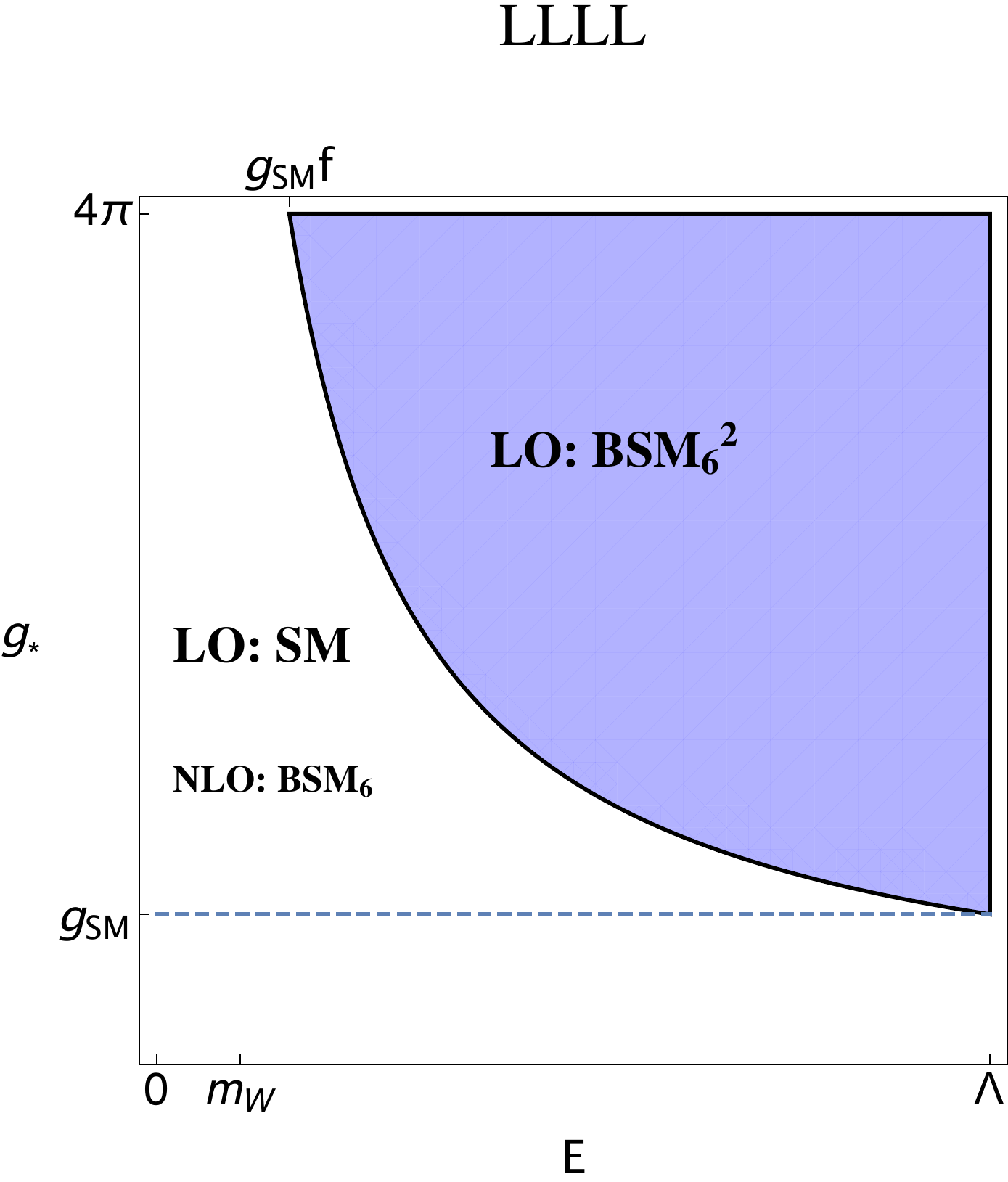}\hspace{2mm}
\includegraphics[width=0.317\textwidth]{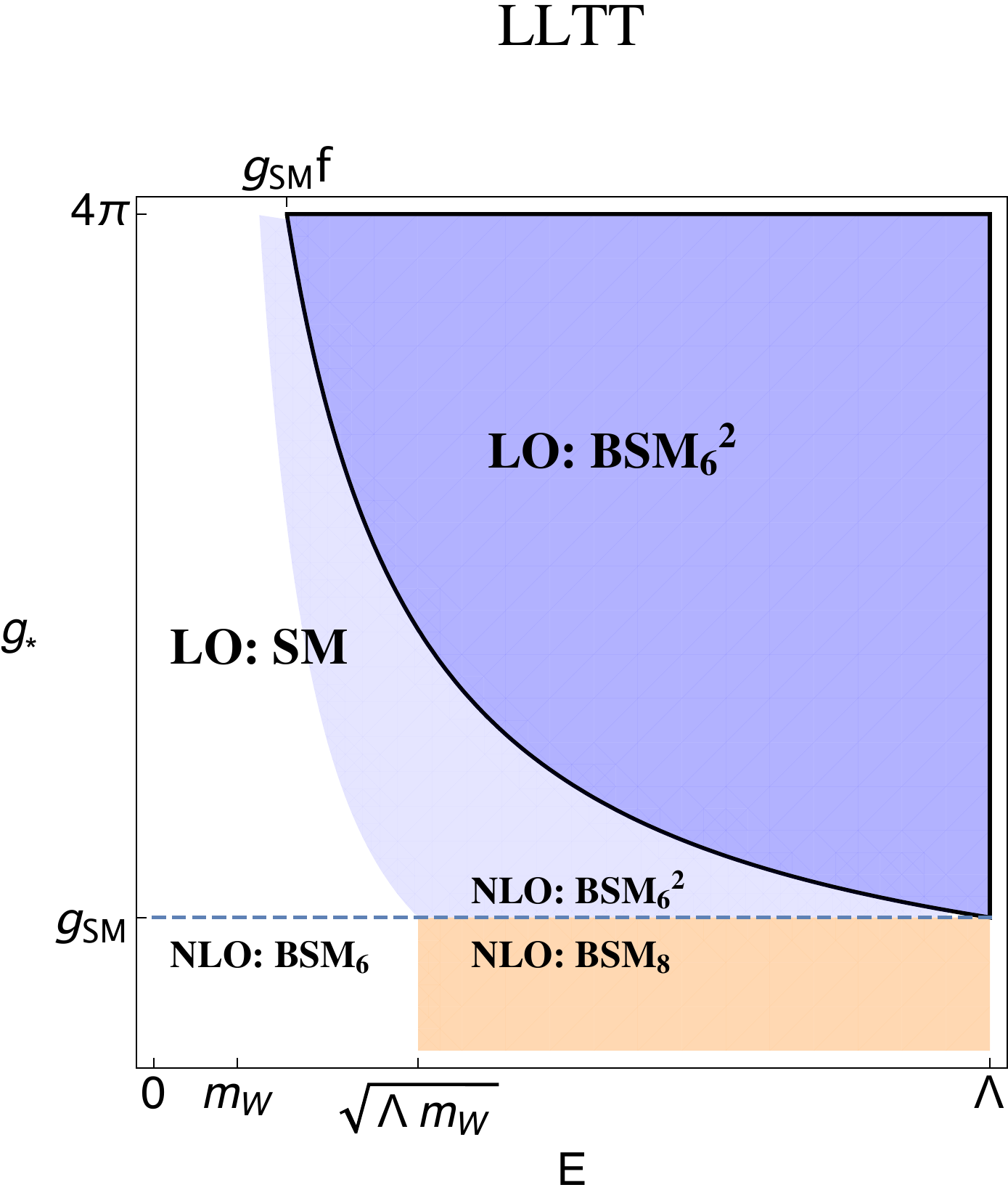}\hspace{2mm}
\includegraphics[width=0.32\textwidth]{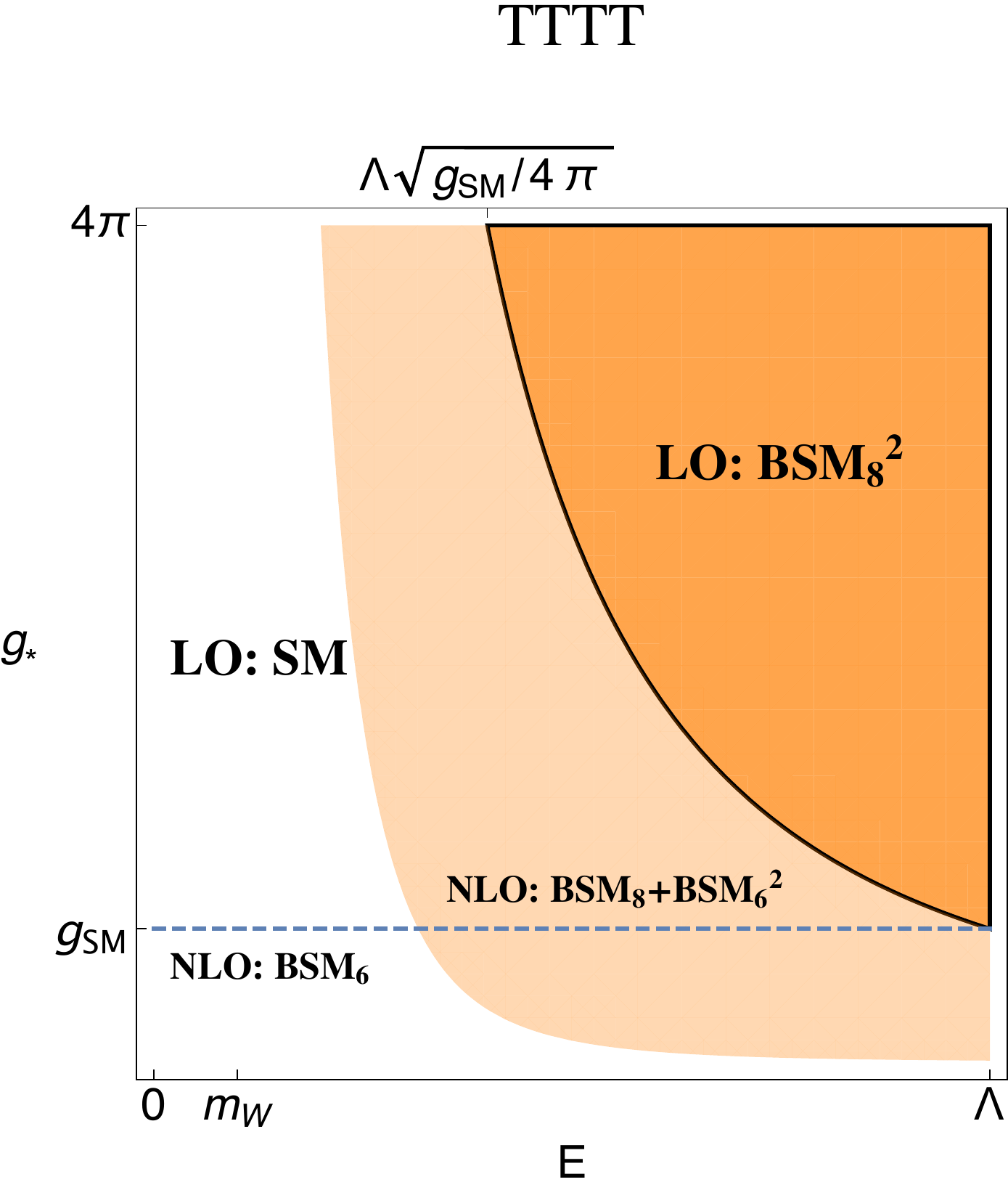}
     \caption{\emph{A schematic representation of the relative size of different contributions to the $VVVV$ scattering cross sections, with polarization 
LLLL (left panel), LLTT (central panel) and TTTT (right panel). LO/NLO denote the leading/next-to-leading contributions to the cross section. In the white region the 
SM dominates and the leading BSM correction comes from the BSM$_6$-SM interference (denoted as $BSM_6$). BSM non-interference is responsible for the light-shaded
 blue and orange  regions, where  the BSM, although it is only a small perturbation around the SM, is dominated by terms of order $E^4/\Lambda^4$, either from 
(BSM$_6$)$^2$ or from the BSM$_8$-SM interference (denoted as $BSM_8$).}}\label{f:plot}
     \end{center}
\end{figure*}

To appreciate the relevance of  non-interference, let us first discuss the BSM amplitudes which do interfere with the SM, such as $A(\phi\phi\phi\phi)$ in the 
scattering of 4 longitudinally-polarized vector bosons. This process receives a contribution from the operator ${\cal O}_H=(1/2) (\partial^\mu |H|^2)^2$, with
estimated coefficient $c_H\approx g_*^2/\Lambda^2$, which may capture for example the effect of Higgs compositeness or the virtual exchange of heavy vectors 
coupled to the Higgs current with strength $g_*$.
At the \deight level, higher-derivative operators will also contribute with estimated coefficients $c^{(8)}\approx g_*^2/\Lambda^4$ (for example they may capture 
higher-order terms in the $p^2/\Lambda^2$ expansion of the propagator of the heavy vectors).
The corresponding contributions to the $V_LV_L\to V_LV_L$ scattering cross section are, schematically,
\begin{equation}\label{LLLL}
\sigma_L \sim \frac{g_{\text{SM}}^4}{E^2}\bigg[ 
 1\,+\!\underbrace{\frac{g_*^2}{g^2_{\text{SM}}} \frac{E^2}{\Lambda^2}}_\text{BSM$_6\times\,$SM}\!
   +\underbrace{\frac{g_*^4}{g^4_{\text{SM}}} \frac{E^4}{\Lambda^4}}_\text{BSM$_6$$^2$} \,
   +\!\underbrace{\frac{g_*^2}{g^2_{\text{SM}}}\frac{E^4}{\Lambda^4}}_\text{BSM$_8\times\,$SM} \! + \,\text{...}\, \bigg]\,.
\end{equation}
Since $E\ll \Lambda$ for the EFT expansion to make sense, \deight effects are always subdominant, while the BSM$_6$-SM interference term  
always dominates for weakly-coupled theories. In the case of strongly-coupled theories, $g_* > g_{SM}$, 
the BSM contribution is  larger than the SM one at energies $E\gtrsim \Lambda \, (g_{SM}/g_*)$, where the $(\text{BSM}_6)^2$ term dominates.
We illustrate this situation in the left panel of Fig.~\ref{f:plot}.
Similar arguments hold for $\psi\psi\to\psi\psi$~\cite{Domenech:2012ai} and $\psi\psi\to\phi\phi$~\cite{Biekoetter:2014jwa}.

As an example where the non-interference is at work, consider the scatterings $V_TV_T\to V_LV_L$ (with its crossings) and $V_TV_T \to V_TV_T$.
We will be inclusive on the transverse polarizations,  implying a sum (or average) over them in the following discussion. We will later
highlight the advantages of an angular distribution analysis able to select the final-state polarizations. 
Let us discuss first the scattering  $V_TV_T\to V_LV_L$. In this case the  largest BSM correction potentially comes from operators of the form $F^2\phi^2$ and $F^3$,
whereas $\phi^4 D^2$ and $\phi^6$ contribute only at subleading level in $\varepsilon_V$.
The helicity selection rules of Section~\ref{sec:helicities} imply that the interference with the SM is suppressed and arises at order $\varepsilon_V^2$ in the
mass insertion or via 1-loop EW corrections. It turns out that the latter effect is always subdominant in the following discussion and will be thus neglected 
for simplicity. The naive estimate of the various terms entering the cross section is different, according to Eq.~(\ref{eq:counting}), for the operators $F^2\phi^2$ 
and $F^3$. Assuming that only $F^2\phi^2$ contributes, one finds, schematically, 
\begin{equation}\label{LLTT}
\begin{split}
\sigma_{LT} \sim \frac{g_{\text{SM}}^4}{E^2}\bigg[ 
 1\, & +\, \overbrace{\frac{g_*^2}{g^2_{\text{SM}}}\frac{m_W^2}{\Lambda^2}}^\text{BSM$_6\times\,$SM}
          +\overbrace{\frac{g^4_*}{g^4_{\text{SM}}}\frac{E^4}{\Lambda^4}}^\text{BSM$_6$$^2$} \\[0.2cm]
      & +\underbrace{\frac{g^2_*}{g^2_{\text{SM}}} \frac{E^4}{\Lambda^4}}_\text{BSM$_8\times\,$SM}+\dots\bigg].
\end{split}
\end{equation}
The importance of the various terms is illustrated in the central panel of Fig.~\ref{f:plot}. For small enough energy, where the BSM gives a small perturbation 
to the SM prediction, the BSM$_6$-SM interference dominates. The suppression of the latter has however an important impact on the behavior at higher energies.
If $g_* > g_{SM}$, it implies a precocious onset of the regime where the $(\text{BSM}_6)^2$ term must be included:
for $ (m_W \Lambda \,g/g_*)^{1/2}<E< \Lambda \,g/g_*$, corresponding to the light blue region of the Figure, the SM still dominates but the $(\text{BSM}_6)^2$ 
term gives the largest correction; for higher energies $(\text{BSM}_6)^2$ eventually dominates the cross section.
For weak or super-weak UV completions, $g_* < g_{SM}$, the largest correction to the SM prediction comes from \deight operators, in particular from
the interference BSM$_8$-SM, as soon as the energy is larger than $\sim\sqrt{m_W\Lambda}$ (light orange region in the Figure).
In this case, an EFT analysis in terms of \dsix operators alone is insufficient.

Yet a different energy behavior is found for the scattering $V_TV_T \to V_TV_T$, where $F^3$ gives the leading correction, while the operators $F^2\phi^2$,
$\phi^4 D^2$ and~$\phi^6$ contribute at sub-leading order in $\varepsilon_V$. (Similar conclusions are in fact obtained also for $V_TV_T \to V_LV_L$ in the case
in which only $F^3$ contributes.) Because the coefficient of $F^3$ scales with only one power of~$g_*$ according to Eq.~(\ref{eq:counting}), the size of the \dsix 
terms (both $(\text{BSM}_6)^2$ and the BSM$_6$-SM interference) is suppressed compared to Eq.~(\ref{LLTT}). The correction from \deight operators might 
not carry a similar suppression, as it happens for example for the $F^2 \bar F^2$ operator, whose  coefficient has a naive estimate 
$c^{(8)} \approx g_*^2/\Lambda^4$.
The different contributions to the cross section can thus be schematically summarized as follows:
\begin{equation}
\begin{split}
\sigma_{T} \sim \frac{g_{\text{SM}}^4}{E^2}\bigg[ 
 1 & +\overbrace{\frac{g_*}{g_{\text{SM}}}\frac{m_W^2}{\Lambda^2}}^\text{BSM$_6\times\,$SM}
       +\overbrace{\frac{g^2_*}{g^2_{\text{SM}}} \frac{E^4}{\Lambda^4}}^\text{BSM$_6$$^2$} \\[0.2cm]
   & +\underbrace{\frac{g^2_*}{g^2_{\text{SM}}}\frac{E^4}{\Lambda^4}}_\text{BSM$_8\times\,$SM} 
       +\underbrace{\frac{g^4_*}{g^4_{\text{SM}}}\frac{E^8}{\Lambda^8}}_\text{BSM$_8$$^2$}+ \dots \bigg]\, .
\end{split}
\end{equation}
Independently of the size of the interference term, this expression shows that as soon as the \dsix effects become bigger than the SM 
(for $E > \Lambda (g/g_*)^{1/2}$),  the \deight contribution takes over and dominates the cross section~\cite{rattetaltoappear}.
Non-interference implies a precocious onset of the regime where \deight operators must be included: for energies $E > \sqrt{m_W\Lambda} \, (g/g_*)^{1/4}$ 
the dominant correction to the SM comes both from $(\text{BSM}_6)^2$ and from the BSM$_8$-SM interference. 
The situation is illustrated in the right panel of Fig.~\ref{f:plot}.
We conclude that, for the scattering $V_TV_T \to V_TV_T$, inclusion of \deight operators is  crucial in a vast energy region above threshold.

So far we have considered processes where the transverse polarizations of the vector bosons are treated inclusively, i.e. they are summed over in the
final state and averaged in the initial one. This in practice corresponded to sum  over two different kinds of helicity amplitudes in each process, one in which 
the SM arises at higher order in $\varepsilon_V$, the other in which it is the BSM$_6$ amplitude to be suppressed. As an example, consider the amplitudes
$A(\phi\phi V^\pm V^\pm)$ and $A(\phi\phi V^\pm V^\mp)$ which have been both included to obtain the estimate of the cross section of $V_LV_L\to V_TV_T$ 
Eq.~(\ref{LLTT}). This suggests that an experimental analysis able to distinguish the polarizations in the final state could be used to select those  processes
where the SM amplitude arises at subleading order in the mass insertion (while the BSM$_6$ contribution is unsuppressed).
This would increase the significance of the signal compared to the irreducible SM background. {Another example is the process} 
$\psi\bar \psi \to V^\pm V^\pm$, relevant for the study of anomalous triple gauge couplings, 
where the polarizations of the final-state vector bosons are equal (while those of the fermions are averaged over). 
In this process 
the leading SM amplitude arises at order~$\varepsilon_V$, while the  BSM$_6$ one is unsuppressed. 
Selecting the final-state polarizations through an angular distribution analysis can thus improve the sensitivity on new physics.
More in general, an exclusive approach to the final state can lead to a parametric enhancement of the signal significance compared to the
naive estimates which follow from our analysis of Section~\ref{sec:helicities}.
An interesting example in this sense is given by the proposal of Ref.~\cite{Dixon:1993xd} to study three-jet events
by exploiting the  distribution of  collinear jet pairs under azimuthal rotations 
as a way to (parametrically) enhance the sensitivity on the operator ${\cal O}_{3G}$.

We conclude our discussion on the  impact of non-interference by noticing
an interesting fact: with the exception of $\psi\psi\psi\psi$,  the SM amplitudes that do interfere with the BSM are accidentally suppressed in their contribution
to inclusive cross sections. Indeed, the contribution of the $V_LV_L\to V_LV_L$ amplitude to the  $VV\to VV$ inclusive cross section  is accidentally suppressed in the 
SM by a factor $\sim 1/500$  with respect to $V_TV_T\to V_TV_T$~\cite{Contino:2010mh}. Similarly, in the SM 
the contribution of $\psi\bar\psi\to V_LV_L$ is only $\sim 1/10$ of the  $\psi\bar\psi\to VV$  total cross section~\cite{rattetalinprep}.
Therefore, despite arising at leading order in the 
high-energy limit,  the SM-BSM$_6$ interference is 
anyway suppressed by the fact that the SM amplitude is small. Since at the LHC current experimental 
studies mostly focus on unpolarized cross sections, this implies an additional obstacle in extracting useful information on \dsix operators through their
interference with the~SM.

\section{Conclusions}\label{sec:conclusions}

In this paper we have shown that in a theory where the SM is extended by \dsix effective operators,  tree-level 4-point amplitudes are subject 
to  helicity selection rules in the massless limit. These forbid the interference between SM and \dsix BSM contributions for all amplitudes 
involving at least one transversely-polarized vector boson. Such non-interference was noticed before in the literature for few specific operators and processes 
(see~\cite{Simmons:1989zs,Cho:1994yu,Dixon:1993xd}).
Our analysis extends the result in a systematic way to all the \dsix operators, identifying the exceptions in which interference can instead arise.
At the phenomenological level, our analysis implies that the  BSM effects that are naively expected to be dominant in an EFT approach, i.e. those captured by the 
interference of \dsix effective operators with the SM,  are suppressed in the high-energy limit.  
The interference only arises at next-to-leading order in an expansion in mass over energy and
in the 1-loop perturbative parameter $\alpha_s/4\pi$ (or $\alpha_{em}/4\pi$ for processes not involving colored particles).
Interestingly, some of the remaining  amplitudes which do feature interference are accidentally small in the SM, implying  anyway a small interference.
This leads to a reduced sensitivity on new physics, especially in the case of analyses that are inclusive on the polarizations of the final-state particles.
Furthermore,  in many cases of interest and in particular when the underlying theory is weakly coupled, a generic EFT analysis in terms of \dsix operators alone 
is insufficient, as \deight ones give an equally large (if not larger) contribution.
\\

\noindent
{\bf Acknowledgements.} We thank Clifford Cheung, Claude Duhr, Joan Elias Miro, Ricardo Monteiro, Yael Shadmi, Giovanni Villadoro and Giulia Zanderighi 
for interesting discussions. 
C.S.M. is supported by the S\~ao Paulo Research Foundation (FAPESP) under grant 2012/21627-9.
The work of R.C. was partly supported by the ERC Advanced Grant No. 267985 \textit{Electroweak Symmetry Breaking, Flavour and Dark Matter: One Solution for 
Three Mysteries (DaMeSyFla)}.

\appendix
\section{The spinor helicity formalism}
\label{a:def}

We summarize here some useful results on the spinor helicity formalism (see Refs.~\cite{Dixon:1996wi,Elvang:2013cua} for a review).
In this approach, the fundamental objects defining the scattering amplitudes are the spinors
$\ar{p}^{\dot a}$ and $|{p}]^{a}$ transforming as $(1/2,0)$ and $(0,1/2)$ under $SU(2)\times SU(2)\simeq SO(3,1)$. They are independent solutions of the massless 
Dirac equation:
\begin{equation}
\begin{aligned}
v_{+}(p) & = (|p\rbrack_{a},0) \hspace{0.5cm} & \bar{u}_{+}(p)&=(\lbrack p |^{a}, 0)  \\[0.1cm]
v_{-}(p) & =(0,|p\rangle^{\dot{a}}) & \bar{u}_{-}(p) &=(0,\langle p|_{\dot{a}})\, ,
\end{aligned}
\end{equation}
where the subscript $\pm$ corresponds to an helicity $h=\pm 1/2$. Dotted and undotted indices are raised/lowered with the 2-index Levi-Civita tensor. 
A $(1/2,1/2)$ Lorentz vector is written in terms of the spinors as $-\pslash=\ar{p}[p|+|p]\al{p}$, while the polarization vectors for spin-1 massless bosons are
\begin{align}
\epsilon^{\mu}_{-}(p;q)=\frac{\langle p|\gamma^{\mu}|q\rbrack}{\sqrt{2}\lbrack q p \rbrack}\,,\quad\epsilon^{\mu}_{+}(p;q)=\frac{\langle q|\gamma^{\mu}|p\rbrack}{\sqrt{2}\langle  q p \rangle}\, ,
\end{align}
where $q$ is a reference vector whose arbitrariness reflects gauge invariance.
The products of angle and square spinors $\langle p q \rangle \equiv \langle p |_{\dot{a}} |q \rangle^{\dot{a}}$ 
and $\lbrack p q \rbrack \equiv \lbrack p |^{a} |q \rbrack_{a}$ satisfy the properties
\begin{equation}
\langle p p\rangle=\langle p q \rbrack =0\,,\quad \langle p q \rangle \lbrack p q \rbrack =2 p \cdot q=(p+q)^2
\end{equation}
for any $p$ and $q$.

In many theories the basic building blocks for all scattering amplitudes are 3-point amplitudes. Momentum conservation  
in the 3-point vertex $(p_1^{\mu}+ p_2^{\mu}+ p_3^{\mu})=0$ and the on-shell condition $p_i^2=0$ imply $p_i \cdot p_j=0$, which in bra-ket notation reads
\begin{align}\label{3partkin}
\langle 1 2 \rangle \lbrack 1 2 \rbrack =0\, ,~~~\langle 2 3 \rangle \lbrack 2 3 \rbrack =0\, ,~~~\langle 3 1\rangle \lbrack 3 1 \rbrack =0\, .
\end{align}
The only non-trivial solutions are: $\langle 1 2 \rangle=\langle 2 3 \rangle=\langle 3 1 \rangle=0$ \emph{or} 
$\lbrack 1 2 \rbrack=\lbrack 2 3 \rbrack=\lbrack 3 1 \rbrack=0$. This means that the 3-particle amplitudes can depend only on square \emph{or} angle brackets, 
never on both. 

Spinors are defined up to a multiplicative factor, referred to as Little group scaling, 
\begin{align}\label{lgs}
|p_i \rangle \rightarrow t_i |p_i \rangle ~~~~\text{and}~~~~|p_i \rbrack \rightarrow t_i^{-1}|p_i \rbrack \, ,
\end{align}
which leaves the momentum $(p_i)_{a\dot{b}}=-|p_i\rbrack_a \langle p_i|_{\dot{b}}$ invariant. Under such transformation the polarization vector of a spin-1 particle
scales as $t_i^{-2h_i}$ if it has helicity $h_i=\pm 1$.  An on-shell tree-level amplitude thus scales as $t^{-2h_i}$ under the rescaling of a particle $i$ with helicity 
$h_i$, and as $t^{-2h}$, with $h=\sum_i h_i$, when all particles are rescaled.
We have seen that the special 3-particle kinematics described below \eq{3partkin} implies that a 3-point amplitude must depend either on square \emph{or} 
angle brackets. Little group scaling and the request of locality then
fix completely the form of the amplitude to be (at tree level)
\begin{equation}\label{e:3amp} 
A_3 =g \begin{cases} 
\langle1 2\rangle^{r_3}\langle2 3\rangle^{r_1}\langle3 1\rangle^{r_2} &\textrm{for}\,\,h(A_3)\leq 0 \\[0.1cm]
[1 2]^{\bar r_3}[2 3]^{\bar r_1}[3 1]^{\bar r_2}&\textrm{for}\,\,h(A_3)\geq 0
\end{cases}
\end{equation}
where $r_1=h_1-h_3-h_2$, $r_2=h_2-h_1-h_3$ and $r_3=h_3-h_2-h_1$,  while $\bar{r}_i=-r_i$. 
From simple dimensional analysis it follows that the total helicity of a 3-point tree-level amplitude, $h(A_3)$, is fixed by the dimensionality of the coupling
constant~$g$; such relation is given by \eq{sumh} in the main text.

Similar arguments applied to $n$-point amplitudes imply that the total helicity $h(A_n)$ satisfies:
\begin{equation}
n - h(A_n) + [g] = even
\end{equation}
where $[g]$ is the sum of the dimensions of the couplings contributing to the amplitude.
For $[g]$ even, in particular,  it follows that $h(A_n)$ has the same parity as $n$.

\section{Supersymmetric Ward Identities}
\label{a:susy}

As long as all up-type or all down-type Yukawa couplings vanish, the SM fields and interactions can be embedded in a supersymmetric Lagrangian with $R$-parity.
When both kinds are non-vanishing, however, holomorphy of the superpotential requires the introduction of an additional Higgs doublet or explicit supersymmetry 
breaking. Most SM tree-level amplitudes (all those not involving simultaneously up- and down-type Yukawas) can thus be written in supersymmetric form.
$R$-parity  implies that no supersymmetric state propagates in the internal lines, so that these amplitudes are effectively supersymmetric. 
This feature is generically lost in BSM$_6$, although some operators can still be uplifted to a supersymmetric form~\cite{Elias-Miro:2014eia}.

Supersymmetry implies important relations between scattering amplitudes~\cite{Grisaru:1976vm} (see~\cite{Dixon:1996wi,Dixon:2004za} for a review). 
Since the supercharge $Q(\xi)=\bar\xi^\alpha Q_\alpha$ annihilates the vacuum for a generic spinor parameter~$\xi$, 
the following Supersymmetric Ward Identities (SWI) hold for amplitudes made of $n$ arbitrary fields~$\Phi_i$:
\begin{equation}\label{SWI}
\begin{split}
0 & =\langle0|[Q,\mathcal{O}_n]|0\rangle \\[0.1cm]
   & =\sum_i \, \langle0|\Phi_1\cdots[Q,\Phi_i]\cdots\Phi_n|0\rangle\, ,
\end{split}
\end{equation}
where ${\cal O}_n \equiv \Phi_1\cdots\Phi_n$.
For a scalar $\phi$ and a Weyl fermion $\psi$ in the same chiral supermultiplet, and a gaugino $\lambda$ and a gauge boson $V$ in the same vector multiplet,
one has
\begin{equation} \label{commapp}
\begin{split}
[Q(\xi),\lambda^+(k)] & =-\theta\langle\xi k\rangle V^+(k)\,, \\[0.05cm]
[Q(\xi),V^-(k)] & =+\theta\langle\xi k\rangle \lambda^-(k) \\[0.15cm]
[Q(\xi),\phi^\dagger(k)] & =-\theta\langle\xi k\rangle \psi^-(k) \\[0.05cm]
[Q(\xi),\psi^-(k)] & =+\theta\langle\xi k\rangle \phi(k) \, .
\end{split}
\end{equation}
Eq.~(\ref{commapp}) holds also for fields with opposite helicity $\pm\to \mp$ provided one replaces $\phi\leftrightarrow\phi^\dagger$ and 
$\langle\xi k\rangle\to-[\xi k]$.

For $n=4$, taking  $\mathcal{O}_4=\lambda_1^+V^+_2V^+_3V^+_4$ in \eq{SWI} gives
\begin{equation} \label{Otype1}
\begin{split}
0 = & \,\langle\xi k_1\rangle A_4(V_1^+V^+_2V^+_3V^+_4) \\
      & +[\xi k_2] A_4(\lambda_1^+\lambda^+_2V^+_3V^+_4) \\
      & +[\xi k_3] A_4(\lambda_1^+V^+_2\lambda^+_3V^+_4) \\
      & +[\xi k_4] A_4(\lambda_1^+V^+_2V^+_3\lambda^+_4)\,.
\end{split}
\end{equation}
Since (supersymmetric-)gauge interactions conserve helicity, amplitudes involving two gauginos with the same helicity and two gauge fields are vanishing
at tree level.
Then \eq{Otype1} implies  $A_4(V_1^+V^+_2V^+_3 V_4^+)=0$. 
Similarly, by taking $\mathcal{O}_4=\lambda_1^+V^-_2V^+_3V^+_4$ in Eq.~(\ref{Otype1}) and choosing
$\xi=k_1$ and $\xi=k_2$, one obtains $A_4(V_1^-V^+_2V^+_3V^+_4)= 0 = A_4(\lambda_1^+\lambda^-_2V^+_3V^+_4)$. 
Finally, $\mathcal{O}_4=\phi_1^\dagger \phi_2 \lambda_3^+ V_4^+$ gives
\begin{equation}
\begin{split}
0 = & \,\langle\xi k_1\rangle A_4(\psi_1^+\phi_2\lambda_3^+V_4^+) \\
      & -[\xi k_2] A_4(\phi_1^\dagger \psi_2^- \lambda_3^+V_4^+) \\
      & +\langle\xi k_3\rangle A_4(\phi_1^\dagger \phi_2 V_3^+V_4^+) \\
      & +[\xi k_4] A_4(\phi_1^\dagger \phi_2 \lambda_3^+ \lambda_4^+)\, .
\end{split}
\end{equation}
The second term in this equation vanishes as a consequence of the $Z_2$ chiral symmetry~(\ref{eq:chiralinv}).
If no cubic scalar term is present in the theory, as we will assume, the same symmetry argument also ensures that the  last amplitude vanishes.
Choosing $\xi=k_1$  then implies $A_4(\phi_1^\dagger \phi_2 V_3^+V_4^+) =0$, while $\xi=k_3$  gives $A_4(\psi_1^+\phi_2\lambda_3^+V_4^+)=0$.

Since supersymmetry commutes with the SM gauge group $G_{\rm SM}$ and the color and Lorentz structures factorize in helicity amplitudes, it follows that
the  above results hold for fermions in generic representations of $G_{\rm SM}$, and not only for gauginos in the adjoint; this proves \eq{MHV}. 
Relations for higher-point amplitudes can be obtained by similar arguments or simply through \eq{sumH}.
Finally, notice that the operators $\psi^2\bar\psi^2$, $\psi\bar\psi\phi^2D$ and $\phi^4D^2$   can be uplift into a supersymmetric  form \cite{Elias-Miro:2014eia},
so that their introduction in the theory will not change the SWI.


\end{document}